\documentclass[submission,copyright,creativecommons]{eptcs}

\usepackage{underscore}
\usepackage[T1]{fontenc}
\usepackage[utf8]{inputenc}

\usepackage{mathtools}
\usepackage{multicol}
\usepackage{pifont}
\usepackage{cite}
\usepackage{tikz}
\usepackage{stmaryrd}
\usepackage{amsthm}
\usepackage{amssymb}
\usepackage{mathrsfs}
\usepackage{varwidth}
\usetikzlibrary{intersections}
\usetikzlibrary{calc}

\DeclareUnicodeCharacter{2218}{$\circ$}
\DeclareUnicodeCharacter{27E6}{$\llbracket$}
\DeclareUnicodeCharacter{27E7}{$\rrbracket$}
\DeclareUnicodeCharacter{220F}{$\Pi$}

\newlength{\contentwidth}
\newcommand{\stepa}{\ding{202}}
\newcommand{\stepb}{\ding{203}}
\newcommand{\stepc}{\ding{204}}
\newcommand{\stepd}{\ding{205}}
\newcommand{\stepe}{\ding{206}}
\newcommand{\stepf}{\ding{207}}
\newcommand{\stepg}{\ding{208}}
\newcommand{\steph}{\ding{209}}

\newcommand{\button}[1]{\texttt{#1}}

\newcommand{\labsec}[1]{\label{sec:#1}}
\newcommand{\refsec}[1]{Section \ref{sec:#1}}
\newcommand{\labfig}[1]{\label{fig:#1}}
\newcommand{\reffig}[1]{Figure \ref{fig:#1}}

\title{A Graphical Interface for Category Theory Proofs in Coq}
\author{Luc Chabassier
  \institute{ENS Paris-Saclay\\ Gif-sur-Yvette, France}
  \institute{Université Paris-Saclay\\ Gif-sur-Yvette, France}
  \institute{INRIA Saclay\\ Palaiseau, France}
  \email{luc.chabassier@inria.fr}
}
\def\titlerunning{A Graphical Interface for Category Theory Proofs in Coq}
\def\authorrunning{L. Chabassier}

\begin{document}
\maketitle

\begin{abstract}
  The importance of category theory in recent developments in both mathematics
  and in computer science cannot be overstated. However, its abstract nature
  makes it difficult to understand at first. Graphical languages have been
  developed to help manage this abstraction, but they have not been used in
  proof assistants, most of which are text-based. We believe that a graphical
  interface for categorical proofs integrated in a generic proof assistant
  would allow students to familiarize themselves with diagrammatic reasoning on
  concrete proofs that they are already familiar with. We present an
  implementation of a Coq plugin that enables both visualization and
  interactions with Coq proofs in a graphical manner.
\end{abstract}

\section{Introduction}
\labsec{intro}

Category theory is an active domain in both computer science and mathematics
research, with applications ranging from algebraic geometry to programming
language design. Its ability to reason abstractly over generic structures has
enabled it to become a unifying language for mathematics, creating bridges
between previously unconnected mathematical
developments~\cite{CaramelloUnification2010}.

The transversal nature of category theory means that once one is familiar with
its vocabulary, it can be used to understand concepts in several other
mathematical theories. However, its abstract nature makes it famously difficult
to understand. A particularly effective tool for managing the abstract nature
of category theory is \emph{diagrammatic reasoning}.

Indeed, many graphical languages have been developed to represent objects and
proofs in different families of categories. Arguably, the most famous of these
is the language of \emph{string diagrams} for monoidal categories, which has
seen applications ranging from physics to logic~\cite{BaezPhysics2010}.
However, the first graphical language usually used when learning categories is
that of \emph{commutative diagrams}, which require only the basic axioms of
category theory.

While domain specific tools already exists for creating and reasoning about
both string diagrams~\cite{BarGlobular2016} and commutative
diagrams,\footnote{\url{https://github.com/varkor/quiver}} they use their own
internal representation of categorical objects, and as such are not linked to
any proof assistant. We propose an approach based on a
Coq\footnote{\url{https://coq.inria.fr/}} plugin\footnote{The plugin can be
found at \url{https://github.com/dwarfmaster/commutative-diagrams}.} that
enables the user to view and progress the proof graphically using commutative
diagrams. Integration in a generic proof assistant means that the user can gain
familiarity with diagram proofs through any specific mathematical domain with
which they are most familiar. Ambroise Lafont's
tool~\cite{LafontAmbroiseYet2023} is another tool to work with commutative
diagrams integrated with Coq with a similar scope to our own. Its development
happened in parallel to ours, and most engineering decisions are different, but
we believe at some point our two tools should converge into one.

Integrating graphical reasoning in proof assistants is not a novel idea, and
previous experiments include Actema~\cite{DonatoDragAndDrop2022} and Lean
widgets~\cite{AyersTool2021}. Lean widgets actually include a specific widget
to visualize the current proof state as a commutative diagram, but is limited
to visualization.

We well begin by demonstrating how our tool is used and what it looks like in
section~\refsec{demo}. Then, we will go into more details into what are the
operations it exposes to the user and how they can be used to progress proofs
in section~\refsec{op}. The next section, section~\refsec{lemmas}, goes into
more details of the lemma applying operation, since a lot of work was needed to
make lemma application graphically intuitive. Finally, in
section~\refsec{architecture}, we detail some of the implementation choices
that we made and the challenges of integrating a graphical interface with Coq.

\section{Demo}
\labsec{demo}

\subsection{Workflow}
\labsec{demo:workflow}

First, we provide an overview of what the tool looks like and how it can be
used. Currently, this plugin works only with the \emph{UniMath} category
library. Of course, porting to another library should be easy enough, but we
have not yet found a satisfactory way to support multiple libraries
simultaneously.

\begin{figure}
  \begin{verbatim}
C : precategory
a, b, c, d : C
m1, m2, m3 : C ⟦ b, c ⟧
m' : C ⟦ a, b ⟧
m'' : C ⟦ c, d ⟧
H1 : m1 = m2 · I
H2 : m3 = m2
f : C ⟦ a, b ⟧ → C ⟦ a, c ⟧
Hf : ∏ m : C ⟦ a, b ⟧, f m = m · m1
============================
I · m' · (m3 · I · (I · m'') · (I · I)) 
  = f m' · (I · (I · I · m'' · I))
  \end{verbatim}
  \caption{A Coq goal}
  \labfig{demo:workflow:goal}
\end{figure}

\reffig{demo:workflow:goal} illustrates the goal considered. It uses the
notation \texttt{C ⟦ a, b ⟧} for the type of morphisms in \texttt{C} from
object \texttt{a} to object \texttt{b}, and \texttt{I} for the identity
morphisms. We see that the context contains some morphisms, some hypotheses
about which morphism is equal to which, and a function \texttt{f} that
constructs new morphisms, along with a hypothesis \texttt{Hf} telling us the
extensionality of this function. The goal itself is quite unclear, being
polluted by an arbitrary association of morphisms, along with many unneeded
identities.

This goal can be normalized by something as simple as:\begin{verbatim}
repeat rewrite assoc.
repeat rewrite id_left.
repeat rewrite id_right.
\end{verbatim}

This Coq script rewrites all associativities to one side, and then removes the
compositions with identities on the left and on the right.

However, we do not need to do this because our plugin has its own normalization
procedure which is applied at the start. More precisely, when started, the
plugin looks into the proof goal to identify the objects and morphisms, and
construct a graph from it. So instead of the previous tactics, we can directly
start the plugin with:\begin{verbatim}
diagram run "test.diag".
\end{verbatim}

The \texttt{"test.diag"} part of the tactic is the file that will be used to
save the interaction and enable the replaying of proofs. This is discussed in
more detail in \refsec{demo:replay}. When this tactic is executed, it starts a
graphical interface, as shown in \reffig{demo:workflow:main}.

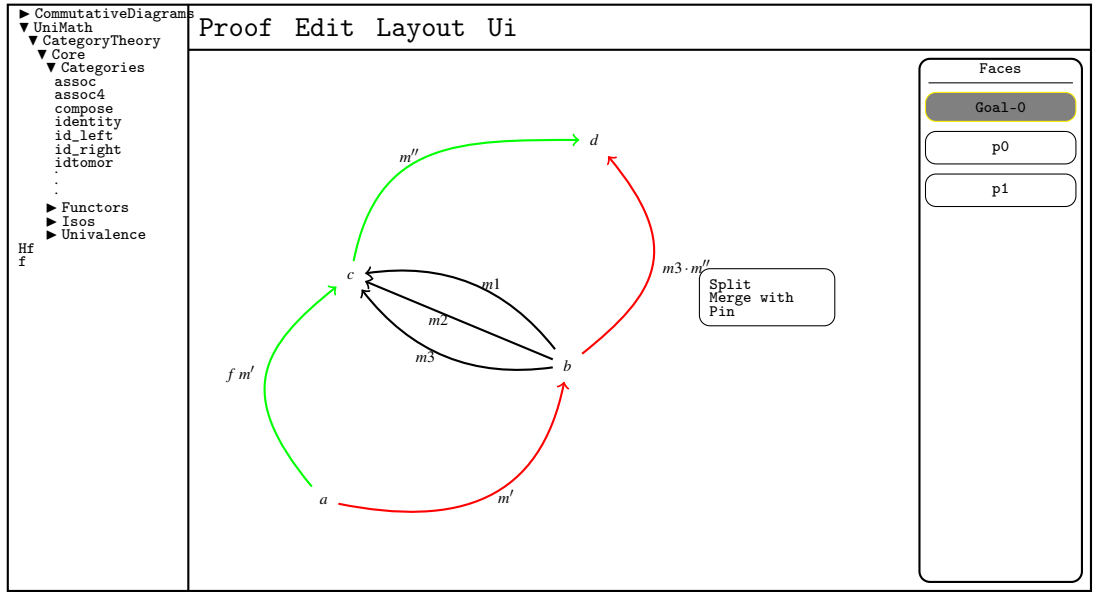
\begin{figure}
  \centering
  \begin{tikzpicture}[thick,font={\tiny},scale=1.2]
  \draw (-2,6.5) rectangle (10,0);
  \draw (0,6.5) -- (0,0);
  \draw (0,6) -- (10,6);
  \draw[rounded corners] (8.1,5.9) rectangle (9.9,0.1);

  \begin{scope}[shift={(-2,0)]},anchor=base west]
    \node at (0,6.35) {$\blacktriangleright$ \texttt{CommutativeDiagrams}};
    \node at (0,6.20) {$\blacktriangledown$ \texttt{UniMath}};
    \node at (0.1,6.05) {$\blacktriangledown$ \texttt{CategoryTheory}};
    \node at (0.2,5.90) {$\blacktriangledown$ \texttt{Core}};
    \node at (0.3,5.75) {$\blacktriangledown$ \texttt{Categories}};
    \node at (0.4,5.60) {\texttt{assoc}};
    \node at (0.4,5.45) {\texttt{assoc4}};
    \node at (0.4,5.30) {\texttt{compose}};
    \node at (0.4,5.15) {\texttt{identity}};
    \node at (0.4,5.00) {\texttt{id\_left}};
    \node at (0.4,4.85) {\texttt{id\_right}};
    \node at (0.4,4.70) {\texttt{idtomor}};
    \node at (0.4,4.40) {\vdots};
    \node at (0.3,4.20) {$\blacktriangleright$ \texttt{Functors}};
    \node at (0.3,4.05) {$\blacktriangleright$ \texttt{Isos}};
    \node at (0.3,3.90) {$\blacktriangleright$ \texttt{Univalence}};
    \node at (0,3.75) {\texttt{Hf}};
    \node at (0,3.60) {\texttt{f}};
  \end{scope}

  \begin{scope}[shift={(0,6.15)},font={\normalsize},anchor=base west]
    \node (a) at (0,0) {\texttt{Proof}};
    \node (b) at (a.east |- 0,0) {\texttt{Edit}};
    \node (c) at (b.east |- 0,0) {\texttt{Layout}};
    \node (d) at (c.east |- 0,0) {\texttt{Ui}};
  \end{scope}

  \begin{scope}[shift={(9,6)},thin,rounded corners,minimum width=2cm,rectangle]
    \node (t) at (0,-0.20) {\texttt{Faces}};
    \draw[thin] (-0.8,0 |- t.south) -- (0.8,0 |- t.south);
    \node[draw=yellow,fill=gray] (goal) at ($ (t.south) + (0,-0.1) $) [anchor=north] {\texttt{Goal-0}};
    \node[draw=black] (p0) at ($ (goal.south) + (0,-0.1) $) [anchor=north] {\texttt{p0}};
    \node[draw=black] (p1) at ($ (p0.south) + (0,-0.1) $) [anchor=north] {\texttt{p1}};
  \end{scope}

  \begin{scope}[shift={(3,3)}]
    \node (a) at (-1.5,-2) {$a$};
    \node (b) at (1.2,-0.5) {$b$};
    \node (c) at (-1.2,0.5) {$c$};
    \node (d) at (1.5,2) {$d$};
    \draw[draw=red,->] (a) .. controls +(1.5,-0.3) and +(-0.3,-1.5) .. (b)
      node[right,midway] {$m'$};
    \draw[draw=green,->] (a) .. controls +(-1,1.2) and +(-1,-0.8) .. (c)
      node[left,midway] {$f~m'$};
    \draw[->] (b) -- (c) node[midway,left] {$m2$};
    \draw (b) edge[->,bend right] node[midway,right] {$m1$} (c);
    \draw (b) edge[->,bend left] node[midway,left] {$m3$} (c);
    \draw[draw=red,->] (b) .. controls +(1,0.8) and +(1,-1.2) .. (d)
      node[midway,right] (clicked) {$m3\cdot m''$};
    \draw[draw=green,->] (c) .. controls +(0.3,1.5) and +(-1.5,0) .. (d)
      node[midway,left] {$m''$};
  \end{scope}

  \begin{scope}[shift={($ (clicked.south east) + (-0.25,0.1) $)},thin,anchor=base west]
    \node (split) at (0,-0.15) {\texttt{Split}};
    \node at (0,-0.30) {\texttt{Merge with}};
    \node (pin) at (0,-0.45) {\texttt{Pin}};
    \draw[rounded corners] (0,0 |- split.north) rectangle (1.5,0 |- pin.south);
  \end{scope}
\end{tikzpicture}
  \caption{The plugin interface}
  \labfig{demo:workflow:main}
\end{figure}

This interface is split into three parts. The left column displays the tree of
lemmas. We will go into more depth about lemmas in \refsec{lemmas}, but as a
first approximation a lemma is a categorical object that is quantified, so you
can see \texttt{f} and \texttt{Hf} in the lemma list. The graph is displayed in
the center. The layout procedure is interactive; therefore different objects
can be dragged around, and the graph will adapt. Furthermore, one can
right-click on different objects to display a contextual menu with available
actions. Finally, the equalities are shown in the right column.

An equality is a Coq term whose type is an equality between two morphisms with
the same source and target. In the graph, the two sides are paths, since the
left and right hand sides of the equality may be composed morphisms. When an
equality is selected, its left-hand side is shown in red, and its right-hand
side in green. We can see that the first equality is selected because its
background is a clearer gray, and looking at the red and green paths in the
graph, we can conclude that it is an equality between \texttt{m' · m3 · m''}
and \texttt{f m' · m''}.

Furthermore, since the equality is surrounded by yellow, the equality is a goal
and not a hypothesis. Now that we can understand the interface, we can see that
what we need to prove is \texttt{m' · m3 · m'' = f m' · m''}. Note that all the
identities are automatically hidden.

The interface exposes some operations to transform the graph and progress the
proof, such as splitting an edge along compositions, splitting an equality
along its planar decomposition, an automatic procedure for faces\dots~These
operations will be detailed in section \refsec{op}. One can progress quite a
lot using only these, but to conclude this proof using \texttt{Hf} is necessary
to deal with \texttt{f}.

\begin{figure}
  \centering
  \begin{tikzpicture}[thick,font={\tiny},scale=1.2]
  \draw (-2,6.5) rectangle (10,0);
  \draw (0,6.5) -- (0,0);
  \draw (0,6) -- (10,6);
  \draw[rounded corners] (8.1,5.9) rectangle (9.9,0.1);

  \begin{scope}[shift={(-2,0)]},anchor=base west]
    \node at (0,6.35) {$\blacktriangleright$ \texttt{CommutativeDiagrams}};
    \node at (0,6.20) {$\blacktriangledown$ \texttt{UniMath}};
    \node at (0.1,6.05) {$\blacktriangledown$ \texttt{CategoryTheory}};
    \node at (0.2,5.90) {$\blacktriangledown$ \texttt{Core}};
    \node at (0.3,5.75) {$\blacktriangledown$ \texttt{Categories}};
    \node at (0.4,5.60) {\texttt{assoc}};
    \node at (0.4,5.45) {\texttt{assoc4}};
    \node at (0.4,5.30) {\texttt{compose}};
    \node at (0.4,5.15) {\texttt{identity}};
    \node at (0.4,5.00) {\texttt{id\_left}};
    \node at (0.4,4.85) {\texttt{id\_right}};
    \node at (0.4,4.70) {\texttt{idtomor}};
    \node at (0.4,4.40) {\vdots};
    \node at (0.3,4.20) {$\blacktriangleright$ \texttt{Functors}};
    \node at (0.3,4.05) {$\blacktriangleright$ \texttt{Isos}};
    \node at (0.3,3.90) {$\blacktriangleright$ \texttt{Univalence}};
    \node at (0,3.75) {\texttt{Hf}};
    \node at (0,3.60) {\texttt{f}};
  \end{scope}

  \begin{scope}[shift={(0,6.15)},font={\normalsize},anchor=base west]
    \node (a) at (0,0) {\texttt{Proof}};
    \node (b) at (a.east |- 0,0) {\texttt{Edit}};
    \node (c) at (b.east |- 0,0) {\texttt{Layout}};
    \node (d) at (c.east |- 0,0) {\texttt{Ui}};
  \end{scope}

  \begin{scope}[shift={(9,6)},thin,rounded corners,minimum width=2cm,rectangle]
    \node (t) at (0,-0.20) {\texttt{Faces}};
    \draw[thin] (-0.8,0 |- t.south) -- (0.8,0 |- t.south);
    \node[draw=yellow] (goal) at ($ (t.south) + (0,-0.1) $) [anchor=north] {\texttt{Goal-0}};
    \node[draw=black] (p0) at ($ (goal.south) + (0,-0.1) $) [anchor=north] {\texttt{p0}};
    \node[draw=black] (p1) at ($ (p0.south) + (0,-0.1) $) [anchor=north] {\texttt{p1}};
  \end{scope}

  \begin{scope}[shift={(5,3)}]
    \node (a) at (-1.5,-2) {$a$};
    \node (b) at (1.2,-0.5) {$b$};
    \node (c) at (-1.2,0.5) {$c$};
    \node (d) at (1.5,2) {$d$};
    \draw[->] (a) .. controls +(1.5,-0.3) and +(-0.3,-1.5) .. (b)
      node[right,midway] {$m'$};
    \draw[purple,->] (a) .. controls +(-1,1.2) and +(-1,-0.8) .. (c)
      node[left,midway] {$f~m'$};
    \draw[->] (b) -- (c) node[midway,left] {$m2$};
    \draw (b) edge[->,orange,bend right] node[midway,right] {$m1$} (c);
    \draw (b) edge[->,bend left] node[midway,left] {$m3$} (c);
    \draw[->] (b) .. controls +(1,0.8) and +(1,-1.2) .. (d)
      node[midway,right] (clicked) {$m3\cdot m''$};
    \draw[->] (c) .. controls +(0.3,1.5) and +(-1.5,0) .. (d)
      node[midway,left] {$m''$};
  \end{scope}

  \begin{scope}[shift={(0.9,4.7)}]
    \draw[rounded corners,fill=white] (-2.8,1.7) rectangle (2.8,-1.7);
    \node[anchor=north] (title) at (0,1.7) {\texttt{Applying Hf}};
    \draw[thin] (-2.7,0 |- title.south) -- (2.7,0 |- title.south);
    \begin{scope}[rectangle,rounded corners,thin,text depth=0.03cm]
      \node[draw,anchor=south east] (apply) at (2.7,-1.6)
        {\texttt{Apply}};
      \node[draw,anchor=base east] (cancel) at ($ (apply.base west) + (-0.1,0) $) 
        {\texttt{Cancel}};
    \end{scope}
    \coordinate (top) at ($ (title.south) + (0,-0.1) $);
    \begin{scope}[shift={(2.20,0 |- top)},thin,rounded corners,minimum width=1cm,rectangle]
      \coordinate (bottom) at ($ (apply.north) + (0,0.1) $);
      \draw (-0.5,0) rectangle (0.5,0 |- bottom);
      \node (faces) at (0,-0.20) {\texttt{Faces}};
      \draw (-0.4,0 |- faces.south) -- (0.4,0 |- faces.south);
      \node[draw=green,anchor=north] at ($ (faces.south) + (0,-0.1) $) {\texttt{p0}};
    \end{scope}
    \node (lema) at (-2.4,0) {$a$};
    \node (lemb) at (0.3,-1) {$b$};
    \node (lemc) at (0.3,1) {$c$};
    \draw (lema) edge[->,bend right] node[below] (fm0) {$f~?m0$} (lemb);
    \draw (lema) edge[->,bend left] node[left] {$?m0$} (lemc);
    \draw (lemb) edge[->,bend right,orange] node[right] {$m1$} (lemc);
    \begin{scope}[shift={($ (fm0.south east) + (-0.1,0.1) $)},thin,anchor=base west]
      \node (match) at (0,-0.15) {\texttt{Match}};
      \node (pin) at (0,-0.30) {\texttt{Pin}};
      \draw[rounded corners,fill=white] (0,0 |- match.north) rectangle (1.5,0 |- pin.south);
      \node at (0,-0.15) {\texttt{Match}};
      \node at (0,-0.30) {\texttt{Pin}};
    \end{scope}
  \end{scope}
\end{tikzpicture}
  \caption{The lemma application window}
  \labfig{demo:workflow:lemma}
\end{figure}
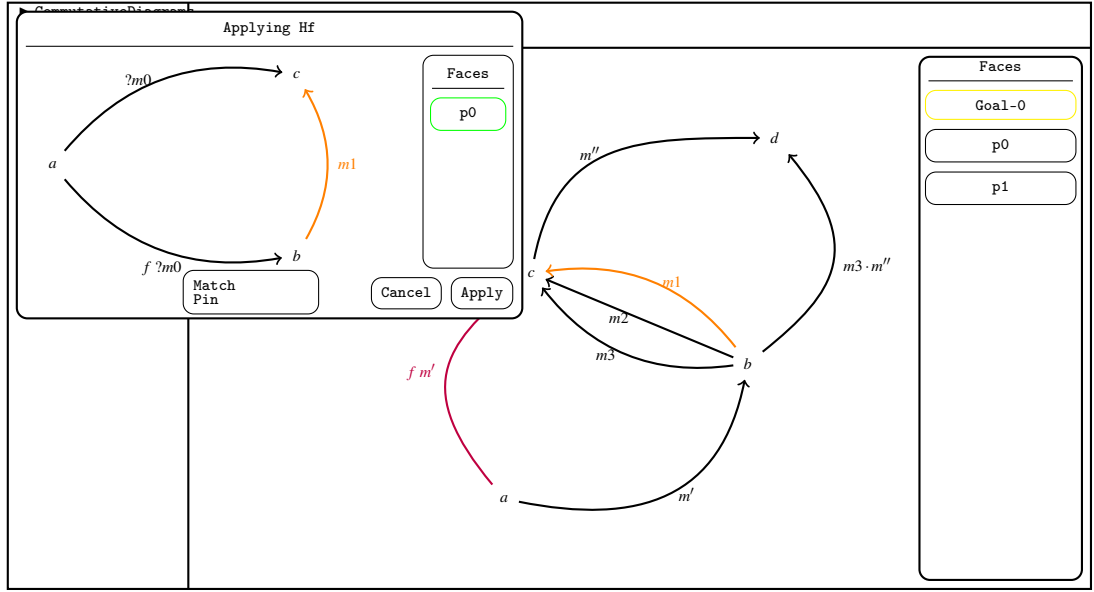

Clicking on \texttt{Hf} opens the lemma application window, as shown in
\reffig{demo:workflow:lemma}. The lemma is displayed as a graph, and the user
can then match objects in the graph lemma with objects in the goal graph.
Matching objects means selecting any object in any window by clicking on it,
and then right-clicking on an object of the same type (node, edge or face) in
the other window and selecting the \emph{match} option.

Matched objects are displayed in orange, and the selected object is shown in
purple. When matching, the underlying Coq terms are unified. Since for
interface Coq terms are opaque objects, the unification is done by Coq itself.
Unification happen every time a pair of objects are matched, not at the end. So
the annotations on the two graphs get progressively more precise as more
objects are matched.

Finally, when the lemma is applied, the pushout of the two graphs modulo the
matching is computed, and the result is the new goal graph. A more precise
description of the lemma application mechanism is provided in \refsec{lemmas}.

Finally, when the user is satisfied with what they have done in the interface,
they can simply click on \button{Finish} in the \button{Proof} menu. The
interface will then close, and the Coq state will be advanced such that any
remaining proof obligation in the interface becomes a new Coq goal.

\subsection{Replaying Proofs}
\labsec{demo:replay}

One thing that is not obvious is how this works when we replay the proof. This
is a common difficulty of graphical approaches because there is no obvious way
to store graphical interactions. The way we decided to do it is that we created
our own very simple language to describe operations on the categorical proof
state. Subsequently, whenever the user interacts with the interface, an
equivalent operation is added to the script. Finally, when the proof is
terminated, the script is saved to a file.

We could have chosen a binary format to save the interactions. However, we
believe that there is value in a simple textual language that can be read,
shared, and written directly. \reffig{demo:replay:example} shows an example of
the script language we are using.

\begin{figure}
  \begin{verbatim}
merge mab mab_0
merge mcd_0 mcd
split mbd
decompose Goal-0 mab:<m3;m2>:mcd;mab:<m2;m1>:mcd
apply Hf b:b a:a mac:mac m1:m1 c:c mab:mab
solve Goal-0-0
succeed
  \end{verbatim}
  \caption{An example of the plugin scripting language}
  \labfig{demo:replay:example}
\end{figure}

There are two variants of the \texttt{diagram} tactic. First, \texttt{diagram
run} tries to run again a previously done proof and only opens the interface in
the first run or if the script fails. Second, \texttt{diagram edit} always
opens the interface, and if a previous proof is found, the interface is set
such that \button{Edit > Redo} replays the different steps one by one, allowing
us to make changes to or visualize the proof.

The question remains as to where to save the proof and where to look it up. We
decided that every time the tactic is called, the user must give a file name as
argument, and the script will be loaded from this file if it exists and saved
to it on success. This approach makes it easy to share the proof script with
other users because it is simply another text file in the Coq development and
introduces no ambiguity as to where the proof comes from. However, this
introduces many small files if the tactic is used in many places, and it makes
reading the proof without executing it slightly more complicated because the
script is in another file.

Another approach is to have the script directly inline in the Coq proof. We
actually tried to implement it for a bit, and got stuck due to limitations in
Coq, but a bigger problem is that there is no way for a Coq plugin to change
the Coq proof. Thus we could either display the script at the end and remind
the user to copy it inside the proof, or perhaps create an integration with
\emph{coq-lsp}.\footnote{\url{https://github.com/ejgallego/coq-lsp}} This is
what Lafont's tool~\cite{LafontAmbroiseYet2023} does, by inserting a Coq
comment containing a single line \emph{JSON} representation of the proof.

Yet another approach, which is the one taken by
Actema~\cite{DonatoDragAndDrop2022}, stores the proofs in a
database\footnote{More precisely, it stores the proof in an \emph{SQLite}
database local to the project.} indexed by a hash of the current proof state.
The main advantage of this approach is that it is completely transparent to
users. However, it is not possible for a human to read the saved proof as it is
stored in a binary format. Furthermore, sharing proofs through git becomes more
complicated, and even a very simple change in the proof state where the tactic
is called can result in Actema not finding the previous proof.

\subsection{Comparison with Other Tools}
\labsec{demo:comparison}

There are many tools to work with different flavors of diagrams; however,
there are mainly three that we are aware of that work directly with commutative
diagrams, in addition to our own.

The first is Quiver.\footnote{\url{https://github.com/varkor/quiver}} Quiver
is an online tool to draw diagrams. As such, the tools it supports and the
range of diagrams it can represent are much greater than what our work can do.
However, it only deals with drawing diagrams and has no logic to perform proofs
with them.

The second is the Lean widget~\cite{AyersTool2021} which displays diagrams.
Similar to our tool, it is integrated into a proof assistant, and can create a
representation of a graph from the proof context. However, it is not
interactive, and the proof cannot be progressed from it. It is only meant as a
visualization helper. Furthermore, it was meant as a demo for the Lean widget
system, but to the best of our knowledge, it is no longer being developed.

The last is Lafont's diagram editor~\cite{LafontAmbroiseYet2023}. It started
as a tool similar to Quiver, more focused on drawing diagrams. However, it has
evolved to become much more similar to our tool, integrating with Coq-lsp to
enable interactive features. As it is now, it has more features to build a
diagram, but the proof generation is less interactive. We expect our tools to
converge on their abilities.

\section{Operations}
\labsec{op}

All provided operations update the \emph{proof state} of the interface in some
way. It includes the Coq proof state as well as some additional state. The
first is the graph. While it is initialized automatically from the Coq state,
it can be modified in ways that do not exactly map to the Coq state, such as
renaming edges or duplicating nodes, and thus must be stored separately.

In addition, there is some additional state that is not \emph{proof-relevant}
but useful to the user nonetheless, that the plugin keeps track of. The main
example is the layout of the graph.

Since goals are simply objects in the graph annotated by existential, the
interface handles multiples goals at once. They are highlighted in yellow to be
made more explicit. However, since the graph is the proof context, it only
handles a single context, so all the goals must have the same context.

\paragraph{Insertion.} The most basic operation involves the insertion of
nodes, edges, or faces. The user must provide a Coq proof term, that will be
parsed by Coq, and used to annotate the new graph object. When inserting, the
plugin tries to be clever, and when adding an edge, it tries to identify the
source and target in the existing graph to add the edge between them. This
operation only modifies the graph and not the Coq proof state.

\paragraph{Merging.} Two nodes, edges or faces can be merged. When doing so,
the graph is modified, but it also unifies the terms annotated on the objects,
which may progress the Coq proof by instantiating existential variables. To
preserve the graph structure, when merging edges/faces, their
source/targets/sides will also be merged.

Since merging objects means unifying the underlying Coq terms, face merging
means unifying the equality terms. In practice it only happens when one of the
two equalities is an existential variable. In this case, unification only
unifies the types of the equalities, and act as a refinement of a goal.

\paragraph{Solving.} We implemented a domain-specific solver that attempts to
construct an equality proof from the other equalities present, using the
structure of the graph. It aims to solve trivial cases that are usually
\emph{left as an exercise for the reader} in paper proofs. This only
instantiates existential variables, and does not modify the graph.

\paragraph{Decomposition} It applies to an equality and considers the layout of
the graph to perform a planar decomposition of the face corresponding to the
equality, and then refines the equality to reduce it to proof obligations for
all the atomic planar regions that compose its face.

\paragraph{Composition} Given a sequence of consecutive edges, one may want to
add an edge in the graph that is annotated by the composition of the annotation
of the previous edges. Those edges are always implicitly present since the
axioms of category theory mandates the existence of the composition of any two
consecutive edges, but it might be interesting to explicit some of them to make
shapes in the diagram more explicit, or to apply lemmas. As the identity is the
special case of the composition of an empty path, this operation also enables
the creation of edges annotated with the identity on a node.

\paragraph{Normalisation} In a way the opposite operation of the previous one.
Given an edge annotated by a composition of morphisms, it splits the edges into
multiple atomic edges. If the edge was annotated by an identity, it is simply
removed.

\paragraph{Lemmas} The user can apply lemmas that \emph{are about categories}.
See \refsec{lemmas} for details on what exactly it means and how precisely it
can be used.

\paragraph{Terminating} The plugin can be terminated at any point during the
interaction. The graph is then forgotten, but the progress made in the Coq
proof state is preserved. This enables going back and forth between the
graphical interface for the proof steps that benefit the most from it, and the
Ltac language for the other steps.

\section{Lemmas}
\labsec{lemmas}

\subsection{Introduction}
\labsec{lemmas:language}

The previously described operations on the state graph allow for most
manipulations that depend only on the categorical structure. However, most
proofs are not in abstract categories, but in either specific categories or
categories with more structure. 

The additional hypotheses that one has on the graph can be used by going back
and forth between Coq and the plugin, performing graph manipulation steps in
the plugin, and then applying lemmas from Coq. This is unsatisfactory for two
reasons. The first one is that, despite our efforts, going back and forth still
necessitates a context switch from the user, which interrupts their flow. The
second is that often lemmas that \emph{talk about categories} are more
naturally thought of in terms of graphs.

Indeed, many properties are defined as shapes of diagrams commutating or
additional constructions that can be made on a diagram. The main examples are
\emph{universal properties}, which are drawn more often than written.

We want to identify a subset of Coq lemmas that we can reasonably represent as
lemmas. We do so by restricting the shape of the lemma statement, and the kind
of object it can mention. We say a lemma \emph{talks about categories} when it
pass our criterion. In \refsec{lemmas:graph}, we describe the creation of an
associated graph for those lemmas. In \refsec{lemmas:application}, we describe
how these graphs can be used to simulate lemma application.

Because Coq logic is powerful, we first restrict ourselves to
\emph{first-order} formulas. More precisely, we consider lemmas whose types are
a sequence of universal and existential quantifications over types that are
\emph{relevant for categories}, and whose conclusion is also \emph{relevant for
categories}. A type is said to be \emph{relevant for categories} if it is the
type of categories, a category, an object, a morphism, a functor, or an
equality between morphisms. Of course, many more types should be supported.
Natural transformations is an obvious one, but we could also support pushouts,
pullbacks, isomorphisms\dots As it is now, support for categorical structures
is hardcoded, so we restricted ourselves to the previous list to limit the
amount of work.\footnote{Making our plugin extensible with regard to structures
is an objective we have, but the work done on this aspect is too preliminary to
be published at the time of writing.}

\subsection{Associated Graph}
\labsec{lemmas:graph}

Once we have a lemma that talks about categories, we want to construct a graph
associated with it. The procedure is the same as that used for extracting the
graph from the proof state, but with a few differences. The first difference is
that, in a quantified formula, the objects under quantifiers depend on the
objects above. Therefore, the first step is to bring all the objects in the
same scope. To do so, we must eliminate the quantifiers.

For universal quantifiers, this is achieved by creating an \emph{existential
variable} of the quantified type and applying the lemma to this variable,
substituting all references to the quantified variable with reference to the
existential variable.

For existential quantifiers, we use the fact that Coq is a constructive
framework in which an existential quantifier is a dependent pair; therefore, we
can substitute references to the quantified variables by the first projection
of the lemma.

By recursively applying these two steps, we obtain a set of terms within the
same scope. We can now apply the same extraction procedure used for the goal to
create the graph associated to the lemma.

In contrast to the graph obtained from the proof state, which has at most one
existential from its conclusion, the graphs extracted from lemmas are filled
with existentials, and most of the terms that are not existentials have some in
them. Therefore, we call the graph associated with a lemma a \emph{pattern}.

\subsection{Application}
\labsec{lemmas:application}

The ability to represent lemmas as graphs is interesting in and of itself as a
visualization aid. However, the notion of applying the lemma is no longer
obvious. We found a way to apply lemmas that use only their graph
representation, which we believe is intuitive. We discuss the expressive power
of such an operation in more detail in \refsec{lemmas:flow}.

The idea is to construct a partial matching between the graph of the lemma and
the graph of the state. Once this matching is built, the \emph{pushout} of the
matching is computed and the result is the new proof state.

Specifically, the user first needs to provide a partial match between the graph
of the lemma and the goal. To do so, they can select objects that must be
identified in the lemma and goal. These objects are interactively unified, so
that a matching is refused as soon as it causes the unification to fail. Once
the user is happy with the match, the interface takes the union of the two
graphs and merges the matched object, which will always succeed because those
objects have already been unified.

\reffig{lemmas:application:prematch} shows such a match given by the user, and
the result is displayed in \reffig{lemmas:application:result}.

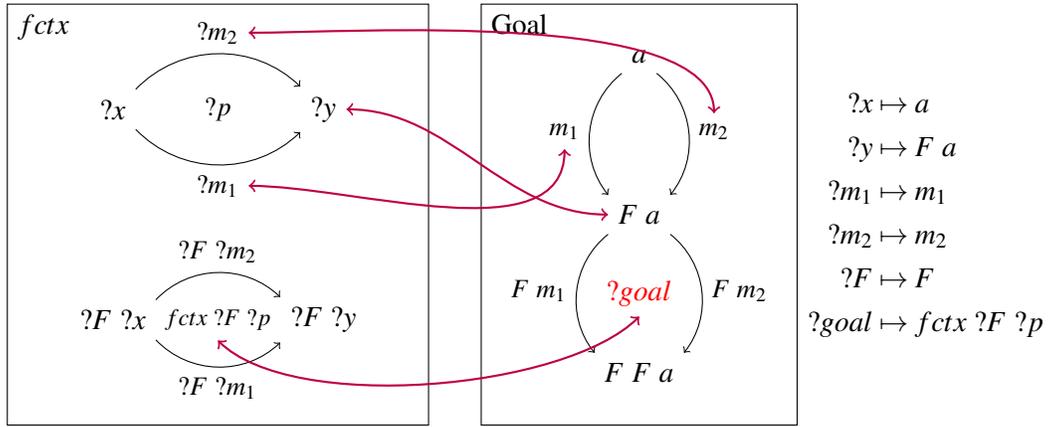
\begin{figure}
  \begin{center}\begin{tikzpicture}[scale=1.4]
    \draw (-1,-1) rectangle (3,3);
    \node[anchor=north west] at (-1,3) {$fctx$};

    \node (lemx) at (0,2) {$?x$};
    \node (lemy) at (2,2) {$?y$};
    \path (lemx.south east) edge[bend right=45, ->]
      node (lemm1) [midway, below] {\small $?m_1$} (lemy.south west);
    \path (lemx.north east) edge[bend left=45, ->]
      node (lemm2) [midway, above] {\small $?m_2$} (lemy.north west);
    \node (p0) at (1,2) {\small $?p$};

    \node (lemfx) at (0,0) {$?F~?x$};
    \node (lemfy) at (2,0) {$?F~?y$};
    \path (lemfx.south east) edge[bend right=45, ->]
      node (lemfm1) [midway, below, text=black] {\small$?F~?m_1$} (lemfy.south west);
    \path (lemfx.north east) edge[bend left=45, ->]
      node (lemfm2) [midway, above, text=black] {\small$?F~?m_2$} (lemfy.north west);
    \node (p1) at (1,0) {\footnotesize $fctx~?F~?p$};

    \draw (3.5,-1) rectangle (6.5,3);
    \node[anchor=north west] at (3.5,3) {Goal};

    \node (a) at (5,2.5) {$a$};
    \node (fa) at (5,1) {$F~a$};
    \node (ffa) at (5,-0.5) {$F~F~a$};
    \path (a.south west) edge[bend right=45, ->]
      node (m1) [midway,left] {\small$m_1$} (fa.north west);
    \path (a.south east) edge[bend left=45, ->]
      node (m2) [midway,right] {\small$m_2$} (fa.north east);
    \path (fa.south west) edge[bend right=45, ->]
      node (fm1) [midway,left] {\small$F~m_1$} (ffa.north west);
    \path (fa.south east) edge[bend left=45, ->]
      node (fm2) [midway,right] {\small$F~m_2$} (ffa.north east);
    \node (goal) at (5, 0.25) {\color{red}$?goal$};

    \draw[<->,thick,purple] (lemy.east) .. controls +(1,0) and +(-1,0) .. (fa.west);
    \draw[<->,thick,purple] (p1.south) .. controls +(0.7,-0.7) and +(-0.7,-0.7) .. (goal.south);
    \draw[<->,thick,purple] (lemm1.east) .. controls +(1,0) and +(0,-1) .. (m1.south);
    \draw[<->,thick,purple] (lemm2.east) .. controls +(1,0) and +(0,1) .. (m2.north);

    \node at (6.5,1) [anchor=west]{$\begin{aligned}
        ?x &\mapsto a \\
        ?y &\mapsto F~a \\
        ?m_1 &\mapsto m_1 \\
        ?m_2 &\mapsto m_2 \\
        ?F &\mapsto F \\
        ?goal &\mapsto fctx~?F~?p \\
    \end{aligned}$};
  \end{tikzpicture}\end{center}
  \caption{A prematch given by the user, with the deduced unifier of the matched terms on the right.}
  \labfig{lemmas:application:prematch}
\end{figure}

\begin{figure}
  \centering
  \begin{tikzpicture}[scale=1.5]
    \node (a) at (0,2.5) {$a$};
    \node (fa) at (0,1) {$F~a$};
    \node (ffa) at (0,-0.5) {$F~F~a$};
    \path (a.south west) edge[bend right=45, ->]
      node (m1) [midway,left] {\small$m_1$} (fa.north west);
    \path (a.south east) edge[bend left=45, ->]
      node (m2) [midway,right] {\small$m_2$} (fa.north east);
    \path (fa.south west) edge[bend right=45, ->]
      node (fm1) [midway,left] {\small$F~m_1$} (ffa.north west);
    \path (fa.south east) edge[bend left=45, ->]
      node (fm2) [midway,right] {\small$F~m_2$} (ffa.north east);
    \node (hyp) at (0, 1.75) {\footnotesize $?p$};
    \node (goal) at (0, 0.25) {\footnotesize $fctx~F~?p$};
  \end{tikzpicture}
  \caption{Result of the application.}
  \labfig{lemmas:application:result}
\end{figure}
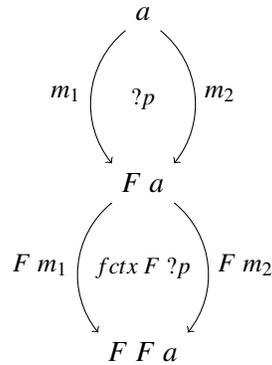

\subsection{Proof Flow}
\labsec{lemmas:flow}

Looking again at the previous example from a higher level, we had a goal we
wanted to prove, and used the lemma to refine it and replace the proof
obligation with the lemma's hypothesis. Thus, this is a case of \emph{backward
reasoning}, which is what someone would do in Coq using the tactic
\texttt{apply} for example.

However, if instead in the state graph there was an equality $p$ between $m_1$
and $m_2$, and nothing between $F~m_1$ and $F~m_2$, the exact same mechanism
would allow to match $p$ with the hypothesis of the lemma, and computing the
pushout would fill the face between $F~m_1$ and $F~m_2$. Again, looking from a
higher level, we applied a lemma to a hypothesis, so this is a case of
\emph{forward reasoning}.

We could also imagine situations where either both the hypothesis and the goal
are present in the state, or neither. The application would still work the in
the same way, and would result respectively in an \texttt{exact} application or
a cut.

We are satisfied with this approach to lemma application, since it allows one
to perform either forward or backward reasoning in a completely uniform way.
Furthermore, we believe that the method is intuitive.

However, it has one major limitation: it cannot apply a lemma under a term.
Indeed, if we have an edge annotated with $f~(m_2 \circ m_1)$, we cannot apply
a lemma under $f$. This is, in general a bigger limitation of our approach, the
fact that it cannot deal with diagrams under terms. To work on this, one would
need in Coq to do a cut to prove that $m_2\circ m_1 = \dots$, and then use the
plugin to prove that cut, so it can still be used, but it is a bit fastidious.

\section{Architecture}
\labsec{architecture}
\subsection{Split Process Organisation}

When we started thinking of writing an interface to work with categorical diagrams,
we did not want to write it in OCaml for two reasons~:\begin{itemize}
  \item Due to difficulties with the linking mechanism in Coq, depending on
    external libraries from a Coq plugin is
    complicated.\footnote{\url{https://github.com/ocaml/dune/issues/5998}}
  \item This would make all the interface development specific to Coq.
\end{itemize}

To circumvent this problem, we designed an abstraction over Coq along with a
protocol to interact with this abstraction, rewrote the plugin to implement
this protocol and expose it locally, and finally implemented the interface in a
separate development in Rust that connects to the plugin and only interacts
with Coq through this protocol.

The final architecture is illustrated in \reffig{architecture:split:overview}.
A successful execution of the plugin proceeds through the following
steps~:\begin{itemize}
  \item[\stepa] The Coq plugin is invoked through the \texttt{diagram run} or
    \texttt{diagram edit} tactics.
  \item[\stepb] It looks at the current goal. If it is an equality between
    morphisms, it constructs the graph from the current proof state,and 
    initializes the model.
  \item[\stepc] It looks for lemmas, and, whenever possible, extracts the graph
    for the lemmas.
  \item[\stepd] It starts the interface as a subprocess, and connects to it
    through its \emph{stdin} and \emph{stdout}.
  \item[\stepe] The interface does a simplification step on the graph.
  \item[\stepf] The interface attempts to replay the previous proof if there is
    one. On success it terminates without opening any windows.
  \item[\stepg] It layouts and displays the graph. At this step, the user can
    interact with the interface and progress the proof using the operations
    described in \refsec{op}.
  \item[\steph] During the interaction, changes have been made to the Coq
    state. Upon success, these changes are used to refine the proof state, and
    the plugin returns to Coq.
\end{itemize}

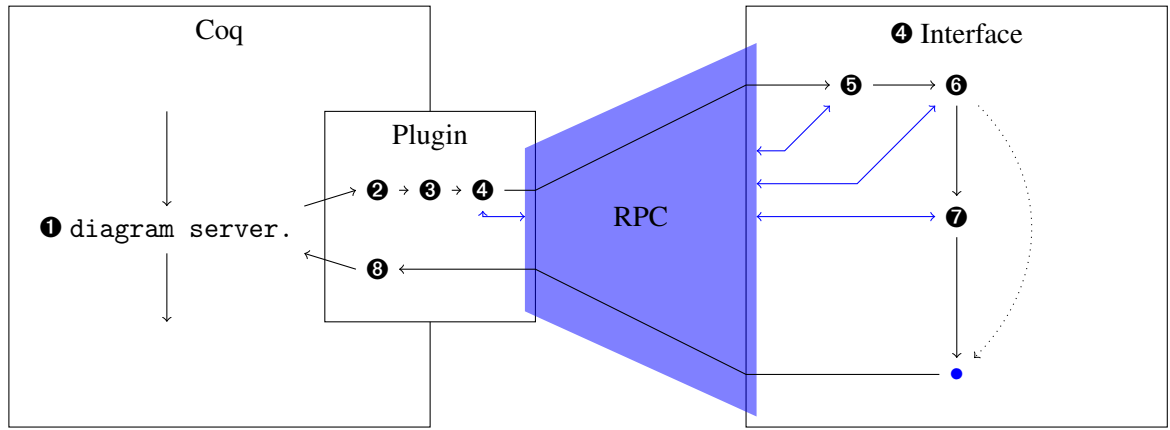
\begin{figure}
  \begin{center}\begin{tikzpicture}[scale=1.4]
    \clip (-0.5,-0.5) rectangle (11.5,4.5);

    \draw (4,3) -- (4,4) -- (0,4) -- (0,0) -- (4,0) -- (4,1);
    \draw (3,3) -- (5,3) -- (5,1) -- (3,1) -- cycle;
    \node at (2,3.75) {Coq};
    \node at (4,2.75) {Plugin};

    \draw (7,0) -- (7,4) -- (11,4) -- (11,0) -- cycle;
    \node at (9,3.75) {\stepd~Interface};

    \fill[fill=blue,fill opacity=0.5] (4.9,1.1) -- (4.9,2.65) -- (7.1,3.65) -- (7.1,0.1) -- cycle;
    \node at (6,2) {RPC};

    \node (tactic) at (1.5,1.875) {\stepa~\texttt{diagram server.}};
    \node (a) at (3.5, 2.25) {\stepb};
    \node (b) at (4, 2.25) {\stepc};
    \node (b2) at (4.5, 2.25) {\stepd};
    \node (c) at (8, 3.25) {\stepe};
    \node (d) at (9, 3.25) {\stepf};
    \node (e) at (9, 2) {\stepg};
    \node (f) at (9, 0.5) {\color{blue}{$\bullet$}};
    \node[coordinate] (g) at (4.5,1.5) {};
    \node (h) at (3.5, 1.5) {\steph};

    \draw[->] (1.5,3) -- (tactic.north);
    \draw[->] (tactic.south) -- (1.5,1);
    \draw[->] (tactic.north east) -- (a.west);
    \draw[->] (a.east) -- (b.west);
    \draw[->] (b.east) -- (b2.west);
    \draw[->] (b2.east) -- (5,2.25) -- (7,3.25) -- (c.west);
    \draw[<->,blue] (b2.south) -- (b2.south |- 4.9,2) -- (4.9,2);
    \draw[->] (c.east) -- (d.west);
    \draw[->] (d.south) -- (e.north);
    \draw[dotted,->] (d.south east) .. controls +(0.7,-0.7) and +(0.7,0.7) .. (f.north east);
    \draw[->] (e.south) -- (f.north);
    \draw[->] (f.west) -- (7,0.5) -- (5,1.5) -- (g) -- (h.east);
    \draw[->] (h.west) -- (tactic.south east);

    \draw[<->,blue] (e.west) -- (7.1,2);
    \path[name path=coriz] (7.1,2.625) -- (11,2.625);
    \path[name path=cdiag] (c.south west) -- +(-5,-5);
    \draw[<->,blue,name intersections={of=coriz and cdiag}] (c.south west) -- (intersection-1) -- (7.1,2.625);
    \path[name path=doriz] (7.1,2.3125) -- (11,2.3125);
    \path[name path=ddiag] (d.south west) -- +(-5,-5);
    \draw[<->,blue,name intersections={of=doriz and ddiag}] (d.south west) -- (intersection-1) -- (7.1,2.3125);
  \end{tikzpicture}\end{center}
  \caption{The architecture of the plugin with the interface}
  \labfig{architecture:split:overview}
\end{figure}

Another advantage of this infrastructure is that, because the Coq plugin
abstracts away the details of Coq, similar plugins implementing the protocol
could be developed for other proof assistants, and the interface could be
reused as-is directly for those other proof assistants. All the complicated
features are implemented in the interface, so it would allow the reuse of a lot
of code with a minimal effort. However, despite our efforts to keep it as small
as possible, the Coq plugin is still around $2000$ \emph{single lines of code},
so not too large but not so trivial to port. We believe that it should be a few
weeks of effort to do something similar for someone used to develop plugins for
another proof assistant.

\subsection{Web Interface}

While the interface has mostly been thought for mathematicians using proof
assistants, we believe it has potential for teaching category theory by hiding
the Coq part. Indeed, we could simply create exercises using Coq, but only
showing the interface to the students so that they could try to do the proof
interactively.

However, installing Coq, a Coq plugin and the interface, which depends on system
libraries for the interface, is complicated. Requiring students to do so would
consume valuable teaching time. To circumvent that, we were asked if it would be
possible to create a web version of our interface. Turns out it is feasible.

The architecture of the web version is a bit peculiar. The web version of the
interface, when started, begins a \emph{websocket} connection to the server. We
implemented a simple websocket server in Rust that starts the Coq process on an
incoming connection, and channels all communication with it through the
websocket. In this way, Coq and the plugin simply need to be installed once on
the server. The architecture is illustrated in \reffig{int:web:architecture}.

One limitation of this approach is that there is no way for the user to
interact with the Coq script. The script must be written on the server, and
only our interface is exposed to the user. However, this is consistent with
the teaching workflow we just described.

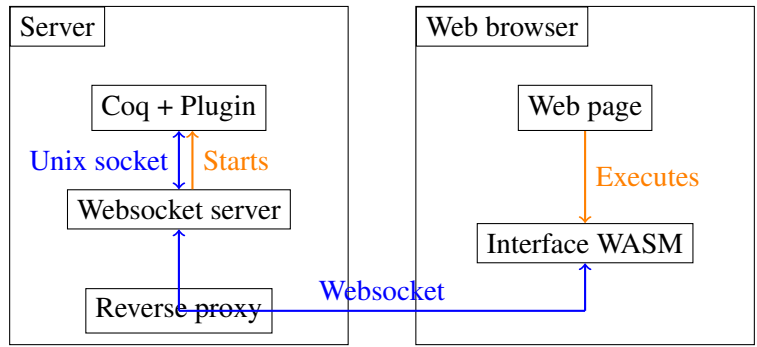
\begin{figure}
  \begin{center}\begin{tikzpicture}[scale=0.9]
    \draw (0,5) rectangle (5,0);
    \node[anchor=north west,draw] at (0,5) {Server};
    \draw (6,5) rectangle (11,0);
    \node[anchor=north west,draw] at (6,5) {Web browser};
    \node[draw] (coq) at (2.5,3.5) {Coq + Plugin};
    \node[draw] (srv) at (2.5,2) {Websocket server};
    \node[draw] (proxy) at (2.5,0.5) {Reverse proxy};
    \node[draw] (int) at (8.5,1.5) {Interface WASM};
    \node[draw] (web) at (8.5,3.5) {Web page};
    \draw (coq) edge[<->,blue,thick] node[left] {Unix socket} (srv);
    \draw (srv) edge[<-,blue,thick] (proxy.center);
    \draw (proxy.center) edge[blue,thick] node[above] {Websocket} (8.5,0.5);
    \draw (8.5,0.5) edge[->,blue,thick]  (int);
    \draw (web) edge[->,orange,thick] node[right] {Executes} (int);
    \draw ($(srv.north) + (0.2,0)$) edge[->,orange,thick] node[right] {Starts} 
          ($(coq.south) + (0.2,0)$);
  \end{tikzpicture}\end{center}
  \caption{Architecture of the web version.}
  \labfig{int:web:architecture}
\end{figure}

We believed that all that was needed was to change the connection method of the
interface to use web sockets. However, we used a synchronous communication
channel, and in the browser it is only possible to use asynchronous
communication. Thankfully, Rust has a good builtin support for asynchronous
coroutines. However, we still need to rewrite much of the interface code in an
asynchronous style. We are in the process of doing so, so we cannot provide a
link to test the web version.

While the motivating aspect of doing this rewrite is to support a web version,
asynchronous communication has other benefits. It makes the interface more
resilient to hangups in the plugin. Indeed, with the current implementation, if
the plugin takes a few seconds to compute something, the interface is frozen
during that time. With the asynchronous version, interaction will be limited
but still possible.

\section{Conclusion}

When doing category theory on paper, the use of diagrams is prevalent. So it
should also be the case when doing category theory proofs with proof
assistants. The fact that one of the first demo for Lean widgets was a widget
displaying the goal as a categorical diagram show the need for such an
interaction. To enable mathematician to do so, we developed a Coq plugin that
automatically constructs a diagram from a Coq context, displays it and allows
the user to progress the proof by graphically manipulating it. Our work is to
be related to Lafont's editor~\cite{LafontAmbroiseYet2023}, another attempt
at creating a graphical way to do category theory proofs.

More generally, the question on how to extend proofs assistants with graphical
representation is still not settled. Both theoretical and practical work has
been done on this subject, but every approach has been different. Indeed, Lean
widgets~\cite{AyersTool2021} use something integrated with the proof assistant
itself, while Actema~\cite{DonatoDragAndDrop2022} needs the interface to be
opened first, and then the plugin connects to it. Our approach of having the
plugin itself start the window is yet another approach. Furthermore, the
question of the replayability of graphical proofs is still open, and in need of
more experiments. We believe our approach of devising a higher-level language
that corresponds to graphical manipulation a good middle ground between text
only proofs, and a fully opaque proof object.

Finally, our development shows the possibility of creating context specific
proof assistants that are layered on top of more general proof assistants. We
believe this to be an interesting direction of research, because it integrates
better the context specific proof assistant in a more general development, and
allows using many such tools together without the need to convert proof from
one to the other. Furthermore, if these tools have good enough abstraction over
the underlying proof assistant, it may reduce the fragmentation of proofs
across different proof assistants, since the proofs in the more specific tool
could be ported from one proof assistant to the other. Indeed, if our own tool
were to be ported to another proof assistant, the scripts in its language could
be executed on the other proof assistant without any changes.

\appendix

\bibliographystyle{eptcs}
\bibliography{thesis}

\newpage
\section{Screenshots}
\labsec{screens}

Since the graphical framework we used only has a black theme, and we did not
take the time to create another lighter theme, we replaced the screenshots in
the main body of the paper by figures made independently. We hope this make
them more readable in the document, but also that it saves ink when printing.
This appendix includes the original screenshots corresponding to the figures

The plugin interface. Corresponds to \reffig{demo:workflow:main}:
\begin{center}
  \includegraphics[width=\contentwidth]{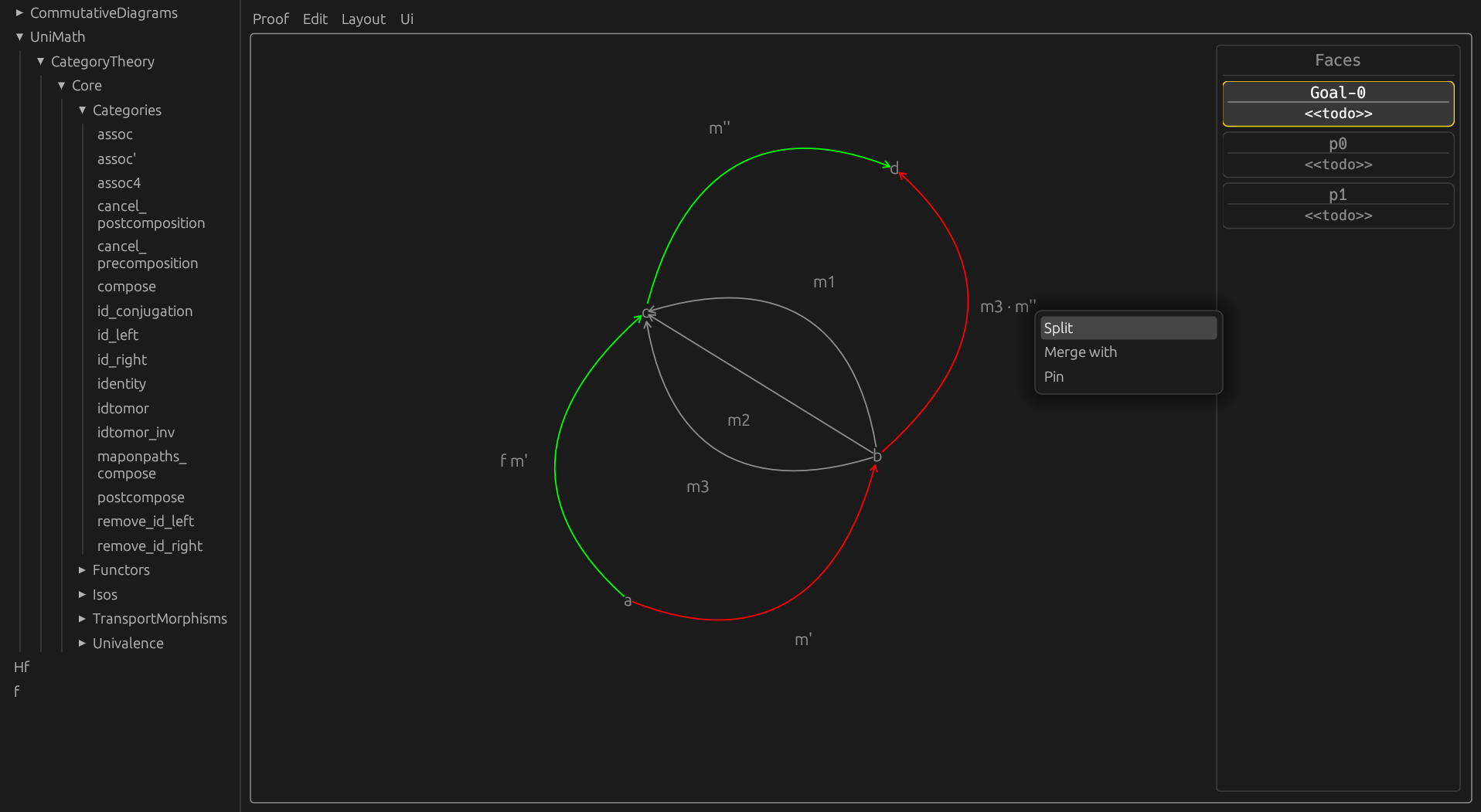}
\end{center}

The lemma application window. Corresponds to \reffig{demo:workflow:lemma}:
\begin{center}
  \includegraphics[width=\contentwidth]{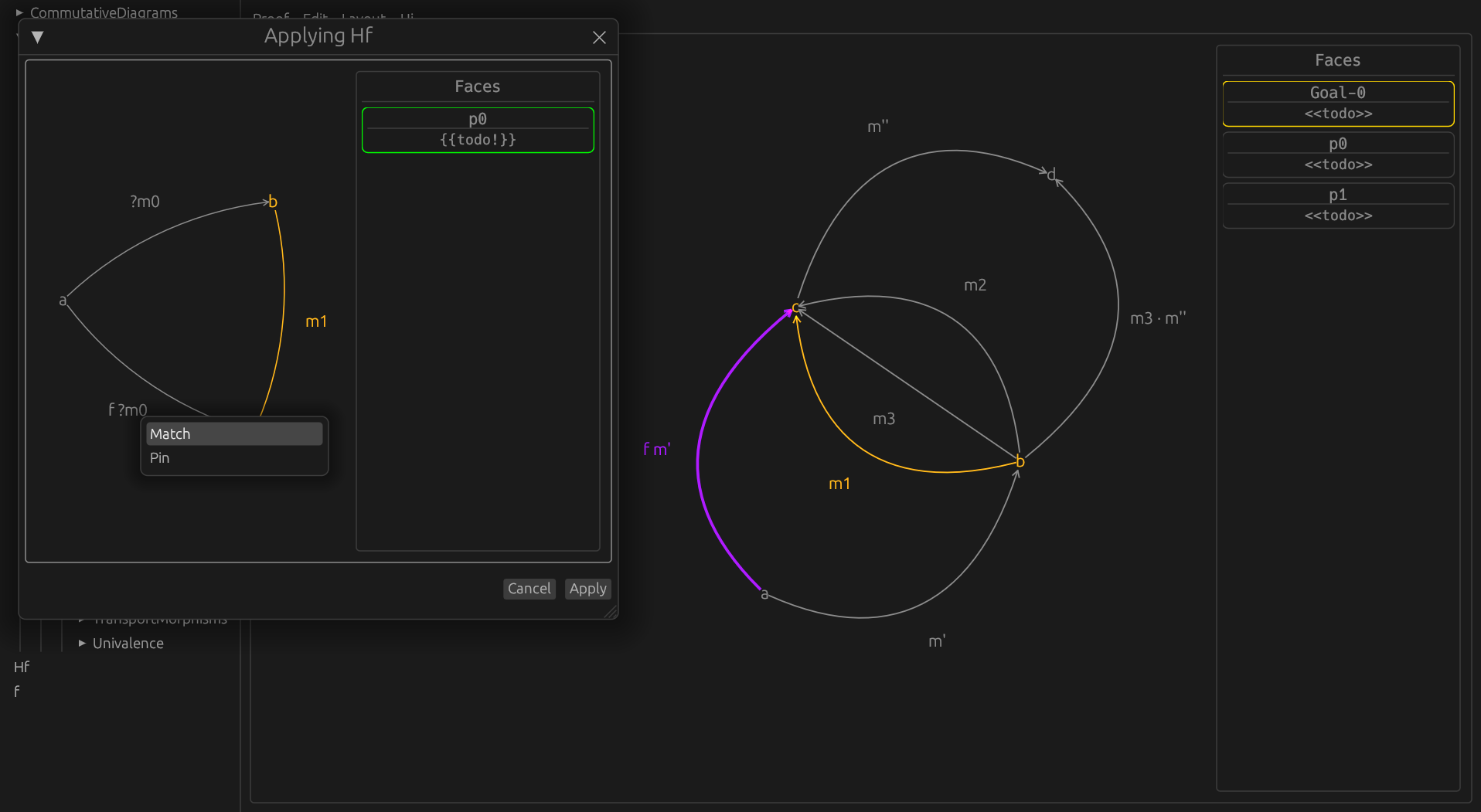}
\end{center}

\end{document}